\documentclass[sigconf]{acmart}

\acmDOI{10.475/1234.5678}

\acmISBN{123-4567-24-567/08/06}

\acmYear{2024}
\copyrightyear{2024}
\usepackage{multirow}
\acmPrice{15.00}

\renewcommand\footnotetextcopyrightpermission[1]{} 

\usepackage{graphicx}
\usepackage{subcaption} 

\begin{document}


\title{LMB: Augmenting PCIe Devices with CXL-\underline{L}inked \underline{M}emory \underline{B}uffer}

\author{Jiapin Wang}
\affiliation{%
  \institution{DapuStor Corporation}
}

\author{Xiangping Zhang}
\affiliation{%
  \institution{DapuStor Corporation}
}

\author{Chenlei Tang}
\affiliation{%
  \institution{DapuStor Corporation}
}

\author{Xiang Chen}
\affiliation{%
  \institution{DapuStor Corporation}
}

\author{Tao Lu}
\affiliation{%
  \institution{DapuStor Corporation}
}




\renewcommand{\shortauthors}{F. Last1 et al.}

\begin{abstract}
PCIe devices, such as SSDs and GPUs, are pivotal in modern data centers, and their value is set to grow amidst the emergence of AI and large models. However, these devices face onboard DRAM shortage issue due to internal space limitation, preventing accommodation of sufficient DRAM modules alongside flash or GPU processing chips. Current solutions either curb device-internal memory usage or supplement slower non-DRAM mediums, prove inadequate or performance-compromising. This paper introduces the Linked Memory Buffer (LMB), a scalable solution utilizing the CXL memory expander to tackle device onboard memory deficiencies. The low-latency of CXL enables LMB to utilize emerging DRAM memory expander to efficiently supplement device onboard DRAM with minimal impact on performance.
\end{abstract}


%
%




\maketitle

\sloppy

\section{Introduction}



\textbf{On-board DRAM shortage in PCIe devices.} Key PCIe devices like SSDs, GPUs, and DPUs in big data retrieval \cite{xdp_2021}, artificial intelligence \cite{G10_2023,bam2023}, and near-data processing \cite{gu2016biscuit, wilkening2021recssd, liang2019cognitive} capitulate to DRAM shortage challenges. 
High-density and large-capacity QLC SSDs \cite{solidigm_d5p5336} are forced to use larger 16KB pages instead of 4KB due to DRAM shortage, suffering increased write amplification. The low indexing efficiency of KV-SSDs \cite{jin2017kaml, kang2019towards, im2020pink} due to lack of memory, hampers their adoption. Memory semantic SSDs \cite{mem_semantic_ssd, CXLssd_2022}, aiming to blend DRAM accessibility and flash durability into a single-tier memory solution, are under development, potentially exacerbating the memory shortage issue.
Insufficient GPU DRAM for large AI models necessitates reliance on SSDs for additional storage \cite{G10_2023,bam2023} at the expense of performance. Likewise, DPUs like XDP \cite{xdp_2021} rely on host memory for substantial index storage, with the on-board memory playing a minor buffer/cache role due to its limitations.

\begin{figure}[tbp]%
    \centering
    {
    \includegraphics[width=6cm]{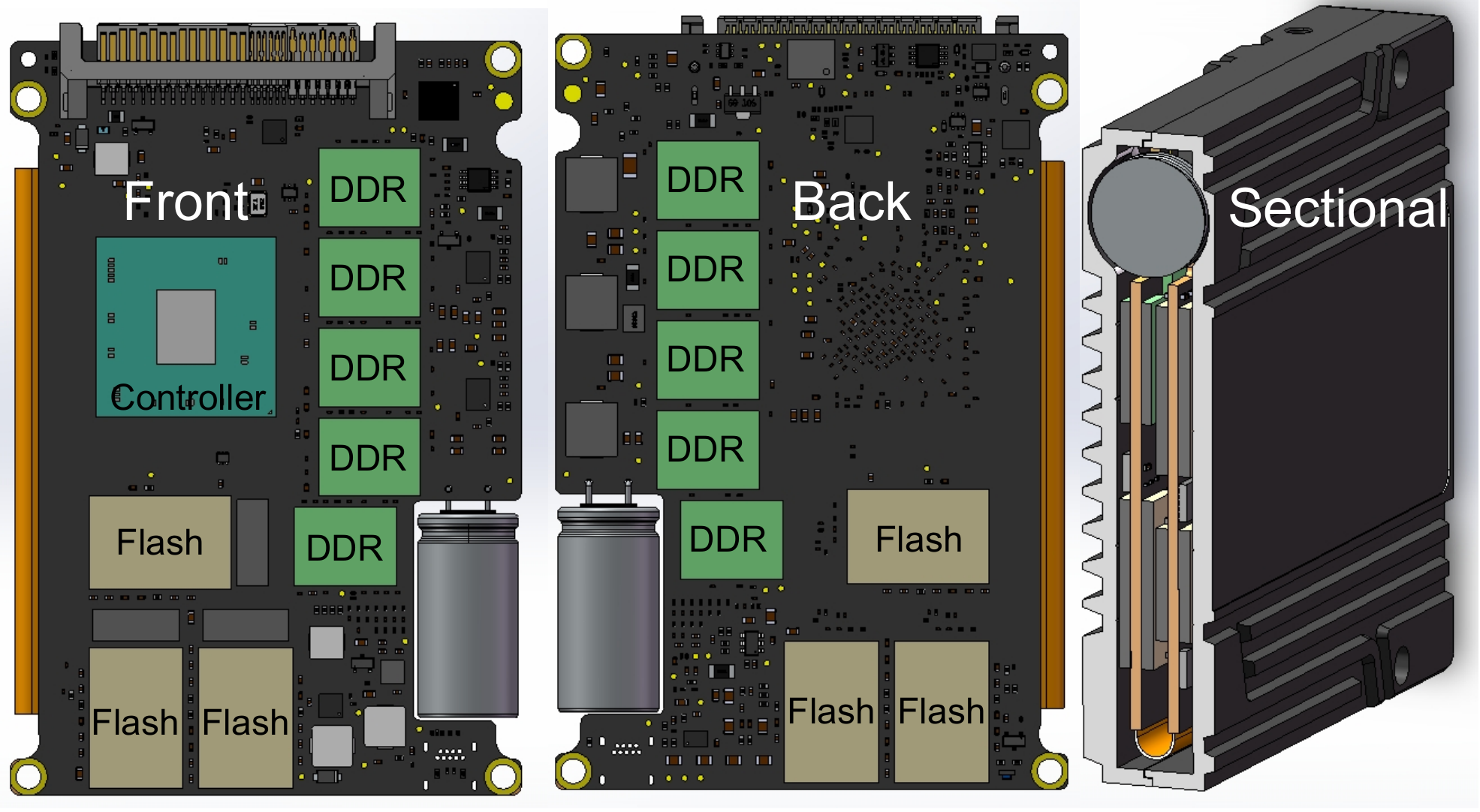}}
    \vspace{-4mm}
    \caption{Internal media layout of a commercial SSD. Obviously, there is no room for more DDR in the ``Inn''.}
\vspace{-4mm}
    \label{ssd_internal_layout}%
\end{figure}

\textbf{Root cause: there is no room for more DRAM in the ``Inn''.}
Enterprise SSD's standard allocation for DRAM is 0.1\% of capacity for 4KB page mapping. Mainstream DRAM technology caps SSD internal memory at 32GB, while QLC technology affords beyond 32TB storage in a U.2 form factor. Nevertheless, integrating additional DRAM in SSDs is a challenge. 
SSD onboard DRAM shortage stems from spatial constraints. DRAM must be situated close to the SSD controller, similar to server memory's proximity to CPU socket, constraining DRAM expansion. Figure \ref{ssd_internal_layout} illustrates the spatial restrictions on the DRAM's positioning in a dual-board SSD, stressing peripheral space limitations near the controller. The growth rate of DRAM density lags behind flash memory, exacerbating the insufficient SSD onboard DRAM issue. Furthermore, DRAM capacity of GPUs is also impeded by spatial restrictions.

\textbf{Existing DRAM conservation and extension solutions possess notable constraints.} Two primary strategies manage internal DRAM shortages: demand suppression and resource supplementation, subdivided into either leveraging internal/external flash or host DRAM. 

DFTL \cite{gupta2009dftl} utilizes flash instead of DRAM for L2P indexing. However, it has performance limitations due to double flash reading, one for index and another for data, thus solely suit for mobile devices. SFTL \cite{jiang2011sftl}/LeaFTL \cite{sun2023leaftl} reduce L2P footprint through index compression, which causes side effects of efficiency and application-specific limitations. DRAM-less host memory buffer (HMB) consumer SSDs underperform in data centers. SSDs adopting large page size diminish index size at the expense of write performance and media lifespan. KV-SSDs mitigate DRAM constraints by sacrificing partial storage or adopting LSM-tree flash index \cite{im2020pink}.

Unified virtual memory \cite{uvm} is a practice to addressing GPU memory shortage, but it causes memory swap overhead. Latest advancements demonstrate high-performance SSDs for memory extension \cite{G10_2023, bam2023}, surpassing CUDA's unified memory in capacity. However, the SSD's slower access speed compared to DRAM creates performance bottlenecks \cite{bam2023}.

\textbf{A universal and scalable solution: linked memory buffer (LMB).} As an SSD supplier, we have faced significant hurdles with memory overhead in large-capacity SSDs. Witnessed the limitations of existing DRAM demand suppression and resource supplementation mechanisms, we believe a system-level architectural solution is needed to address PCIe devices' memory shortage issue.
Driven by the advent of the CXL high-speed interconnection protocol, LMB uses the principle of exchanging time for space to combat device space scarcity. It dynamically expands PCIe device memory, permitting memory sharing between CXL and PCIe devices based on efficient P2P access or host forwarding.

\textbf{LMB challenges.}
Significant issues include dynamic memory allocation, resource optimization, shared resource isolation, access control, cross-device migration, and data security to prevent single points of failure. The dilemma lies between pre-reserving for guaranteed availability or allocating on-demand for efficiency. A single failure in the memory expander can render all devices unavailable. Performance interference due to multiple devices accessing shared memory adds complexity. Addressing these issues is critical.

\textbf{Our vision.}
We propose LMB as a CXL-based memory extension framework and kernel module, using a CXL memory expander as the physical DRAM source. It aims to provide a unified memory allocation interface for PCIe and CXL devices. This permits NVMe and CUDA's unified memory kernel driver to access the CXL memory expander directly and efficiently, enabling SSD and GPU devices to utilize LMB's memory resources as effortlessly as on-board memory.

\section{Background and Related Work}

\subsection{Addressing SSD DRAM Shortage}

\textbf{Flash-backed secondary index.} One way to override the memory wall is to swap indexes to NAND flash. Demand-based FTL(DFTL) \cite{gupta2009dftl} cache frequently used L2P entries in DRAM while retaining the remaining indexes in flash. Spatial-locality-aware FTL(SFTL) \cite{jiang2011sftl} employs spatial locality with strictly sequential access to compress L2P entries. These methods are prone to performance degradation due to the latency discrepancy between flash memory and DRAM.

\textbf{Supplementing device memory using host memory buffer (HMB).} The NVMe 1.2 standard \cite{nvme1_2} enhanced SSD memory using host's DRAM through PCIe's DMA abilities. To offset performance degradation caused by limited SRAM in flash modules for L2P indexing in mobile applications \cite{kim2023integrated}, Host Performance Booster (HPB) places L2P entries in host memory. The HMB scheme challenges the host memory scalability and thus only applicable in the scenario that the DRAM requirement is not large (e.g. hundreds of MBs).

\textbf{Index footprint reduction.} 
As high-density QLC SSD products permeate the market, the need for expanded onboard DRAM for refined L2P mapping tables becomes apparent. The Solidigm D5-p5336 QLC SSD \cite{solidigm_d5p5336} addresses this by introducing a 16KB Indirection Unit (IU) coarse-grained mapping table that reduces DRAM consumption. Although cost-effective, coarse-graining yields issues such as read-modify-write and write amplification compromising SSD's performance. As an alternative, LeaFTL \cite{sun2023leaftl} uses piecewise linear regression for LPA-PPA mappings to minimize index memory overhead. However, it struggles with random writes and DRAM reduction cannot be assured. 

\textbf{Key-value and Memory-semantic SSDs.}
Initiatives like KV-SSD optimize system performance and software efficiency \cite{jin2017kaml, kang2019towards, im2020pink, kwon2022vigil}, but face challenges from extensive DRAM overheads. Memory-semantic SSDs \cite{mem_semantic_ssd, CXLssd_2022} combine cost-effective, byte-addressable DRAM with SSDs for extensive flash space, using memory as CPU-accessible cache. However, they are reliant on DRAM size and cache hit ratios, with misses leading to latency issues and the spatial limitation persists due to the identical form factor as SSDs.

\subsection{Addressing GPU DRAM Shortage}
Deep Neural Networks (DNN) and large-scale model training on GPUs often encounter memory limitations \cite{cano2018survey,li2016performance,wang2022merlin, kwon2018beyond, peng2018optimus}. Mitigation strategies primarily engage unified GPU and host memory \cite{allen2021depth} or SSD-enhanced GPU memory \cite{G10_2023, bam2023}.

\textbf{CUDA unified memory.} Larger host system memory can partially supplement GPU memory constraints. This transparent cushioning is achieved via Unified Virtual Memory (UVM) \cite{uvm,allen2021depth} for flexible data migration and accessibility. However, UVM proves inadequate in obviating GPU memory shortages for extensive dataset training (e.g. LLM) and substantial host-GPU memory migration overhead.

\textbf{SSD-GPU DRAM extension.} BaM \cite{bam2023} and G10 \cite{G10_2023} successfully overcome UVM's data transfer concerns by integrating SSDs as external memory. BaM bypasses intermediary data copying by facilitating direct SSD access to GPU threads. Aside from software and driver changes needed to shift NVMe queues, and IO buffers to GPU memory, BaM demonstrates a marked decrease in SSD-to-GPU data load times during benchmark testing. However, its utility is less perceptible when the graph-processing timeframe exceeds data loading in complex image analysis tasks.

\subsection{Promises of CXL}
\textbf{CXL was proposed to address the scalability constraints of host memory.} 
CXL constructs disaggregated memory through PCIe to overcome the limitations in CPU memory channels and DIMM slots. Hence, we advocate the use of CXL's scalable and disaggregated memory to extend both NVMe HMB \cite{nvme1_2} and Unified Memory infrastructures \cite{uvm}.

\begin{figure}[t]%
    \centering
    {
    \includegraphics[width=8cm]{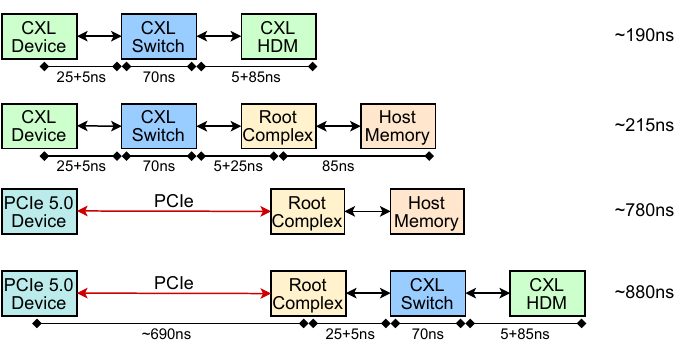}}
    \vspace{-4mm}
    \caption{Estimated latency of PCIe Gen5, and CXL devices accessing host and CXL HDM memory \cite{sharma2022compute,li2023pond}.}%
    \vspace{-4mm}
    \label{fig:latency}%
\end{figure}

\textbf{CXL enables efficient interconnectivity for both Host-to-Device and Device-to-Device.}
By acquiring a Port Based Routing (PBR) ID from connecting either a host or devices to CXL-based PBR switch's Edge Port, it permits access to Global Fabric-Attached Memory (GFAM) devices \cite{CXL_spec}. Exceeding host memory, GFAM's expansibility allows usage within device memory mitigating host applications interference and promoting device-driven enhancements. The Direct P2P attribute bolsters CXL devices in bypassing the host to access similar devices via a shortcut through a switch. CXL port latency is 25ns \cite{sharma2022compute}, 70ns for switch latency including HDM access \cite{li2023pond}, and 780ns for PCIe 5.0 devices accessing host memory, as presented in Figure \ref{fig:latency}. CXL could prove instrumental in surpassing the memory hurdle.

\section{Design of LMB framework}
\begin{table}[t]
\footnotesize
\centering
\caption{CXL and LMB Terminology.}
\vspace{-4mm}
\label{tab:terminology}
\begin{tabular}{p{0.8cm}|p{6.2cm}}
\hline
\textbf{Term} & \textbf{Description}  \\
\hline
\hline
HDM & Host-managed Device Memory \\
\hline
FAM & Fabric-Attached Memory. HDM within a Type 2/3 device that is accessible to multiple hosts. \\
\hline
GFD & Global FAM device \\
\hline
FM &  Fabric Manager controls aspects of the system related to binding and management of pooled ports and devices.  \\
\hline
DPA & Device Physical Address \\
\hline
DMP & Device Media Partition. DPA ranges with certain attributes. \\
\hline
PBR &   Port Based Routing \\
\hline
SPID &  Source PBR ID \\
\hline
SAT & SPID Access Table \\
\hline
\end{tabular}
\end{table}

\begin{figure}[tbp]%
    \centering
    {\includegraphics[width=8cm]{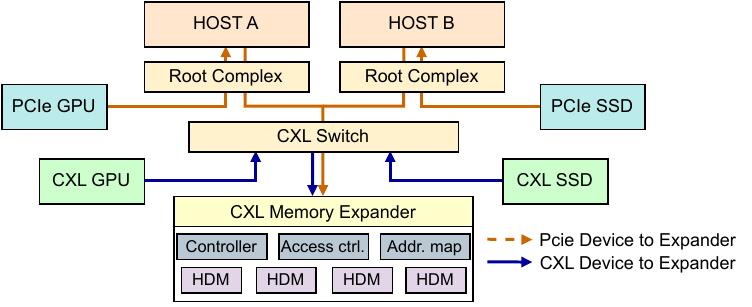}}
    \vspace{-4mm}
    \caption{Overall architecture of LMB.}%
\vspace{-4mm}
    \label{architecture}%
\end{figure}

Building CXL memory pool to extend host memory has been widely studied \cite{gouk2023memory,ha2023dynamic,li2023pond,kim2023smt}, but there lacks attention to extend device memory via CXL. We propose a unified framework called CXL Linked Memory Buffer(LMB) to extend device memory. The overall architecture of the LMB is shown in Figure \ref{architecture} and relevant terminology is listed in Table \ref{tab:terminology}. CXL memory expander is connected to CXL switch, which exposes multiple pooled HDMs to hosts and devices, while providing basic address mapping, access control and other mechanisms. Based on the CXL protocol \cite{CXL_spec}, LMB mounts the Expander as a GFD that can provide memory expansion for either CXL devices attached in the CXL switch or other PCIe devices of the host. More specifically, we plan to implement a unified framework in the kernel that can support both CXL devices and PCIe devices, and provide APIs for device drivers, such as memory allocation and deallocation, access control, etc. For a CXL device, it can directly P2P access the Expander via CXL.mem or UIO access via CXL.io. For a PCIe device, its memory access requests are first converted to CXL.mem requests by host, and then redirected to the Expander.

\subsection{LMB Component}
\begin{figure}[tbp]%
    \centering
    {\includegraphics[width=4.5cm]{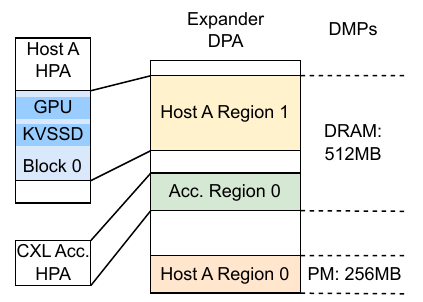}}
    \vspace{-2mm}
    \caption{Expander address mapping.}%
\vspace{-4mm}
    \label{address}%
\end{figure}

\textbf{CXL memory expander and fabric manager.} The Expander serves as a GFD to provide global memory resources to hosts and CXL/PCIe devices in the entire Fabric, supporting large-scale memory pooling and dynamic capacity management. It is directly managed by the FM (Fabric Manager). The FM, a key component in the CXL architecture, is responsible for managing and configuring devices and resources in the CXL Fabric, which can be implemented as software in the host or firmware on a switch or a CXL device. Hosts on Fabric can query and configure the state of Expander through FM APIs to realize dynamic memory allocation among multiple hosts. The Expander manages the internal HDMs to translate HPAs in host requests into DPAs through address mapping. The Expander's DPA space is organized according to Device Media Partition (DMP) and supports DRAM and PM heterogeneous media, as shown in Figure \ref{address}.

\textbf{LMB kernel module.} The memory access protocols of PCIe devices and CXL devices are completely different. Moreover, existing CXL memory pool design is difficult to accommodate PCIe devices. To address this problem, we treat the host as a bridge, and implement the LMB kernel module to provide a uniform memory allocation and sharing interface to both PCIe devices and CXL devices. The kernel module first requests a memory block from the FM and then interacts with the device driver to allocate memory for it. In addition, we promote the loading priority of the LMB module to avoid memory request failure in the initialization phase of the PCIe device driver.

\subsection{Memory Management}
\textbf{Memory allocator.} The kernel module requests memory from the Expander through the FM API \cite{CXL_spec}, and the obtained memory is mapped into the physical address space of the host, waiting to be allocated to the local device. When a kernel module does not have enough free memory to complete the allocation, it requests a single 256MB block from the Expander. When all device memory in a memory block has been freed, the kernel module releases the area to FM. We keep the memory allocator metadata in the host to facilitate the alignment of large memory mappings and avoid triggering multiple CXL memory accesses.

\textbf{Data path.} The PCIe devices cannot directly access the Expander using the CXL protocol, but they can initiate reads and writes to the HDM mapped physical address. The PCIe TLP (Transaction Layer Packet) is converted by the CPU into MemRd/MemWr commands in the CXL.mem protocol. To the Expander, this will be treated as a memory access by the host. Since the PCIe device does not support CXL cache coherency, it sets this memory to the \textit{uncached} type. When the PCIe device and the CXL device share memory, although the PCIe device cannot receive Back-Invalidate Snoop from the Expander, it does not cause consistency problems.

\begin{table}
\footnotesize
\caption{LMB kernel API}
\label{tab:api}
\vspace{-4mm}
\begin{tabular}{p{1.5cm}|p{6cm}}
\hline
\multicolumn{1}{l|}{\textbf{Operation}} & \multicolumn{1}{c}{\textbf{Interface}}                   \\ 
\hline
\hline
\multirow{2}{*}{Allocate}          & lmb\_PCIe\_alloc(*dev, size, *hpa, *mmid)        \\ \cline{2-2} 
                                & lmb\_CXL\_alloc(*CXLd, size, *hpa, *DPID, *mmid) \\ \hline
\multirow{2}{*}{Free}           & lmb\_PCIe\_free(*dev, mmid)                      \\ \cline{2-2} 
                                & lmb\_CXL\_free(*CXLd, mmid)                      \\ \hline
\multirow{2}{*}{Share}          & lmb\_PCIe\_share(*dev, mmid, *hpa)               \\ \cline{2-2} 
                                & lmb\_CXL\_share(*CXLd, mmid, *hpa, *DPID)        \\ \hline
\end{tabular}
\vspace{-4mm}
\end{table}

\subsection{LMB APIs and Access Control}
\begin{figure}[tbp]%
    \centering
    {\includegraphics[width=6cm]{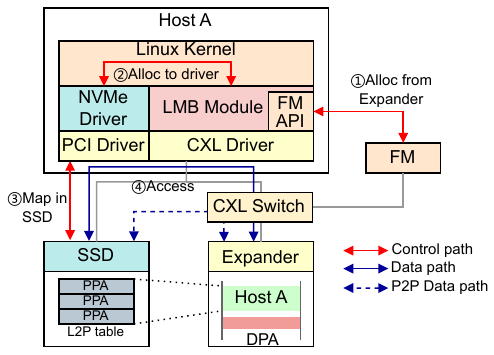}}
    \vspace{-2mm}
    \caption{SSD stores the L2P table through the LMB.}%
\vspace{-4mm}
    \label{ssd_l2p}%
\end{figure}

\textbf{APIs for driver usage.} The kernel module exposes the APIs shown in Table \ref{tab:api} for device drivers that support the \textit{alloc}, \textit{free}, and \textit{share} interfaces. The Figure \ref{ssd_l2p} shows the process of an SSD applying for LMB to store the L2P table.
The PCIe device will obtain the bus address accessible to the device and a unique memory ID in the local host. In addition to the HPA (Host Physical Address) of the GFAM, the CXL device also gets the global PID of the expander to initiate P2P requests. The shared memory can be used as an input or output buffer to reduce memory copy between devices. For example, sending data from SSDs to accelerators for computation, a zero-copy data path can be achieved with the help of shared memory. The kernel module maintains the mapping of HPA and bus addresses to physical addresses, and the memory sharing between PCIe devices and CXL devices can be completed by address translation. 

To protect memory safety between devices, it is necessary to restrict a device from accessing memory ranges that do not belong to it. For PCIe devices, it is common to use IOMMU to isolate the range of memory that can be accessed by the device. The access control of CXL device to GFD is managed by SAT (SPID Access Table), and GFD can identify the CXL device or host that initiates the request according to the SPID field in the memory request. LMB integrates the above two access control modes. When memory is requested by a PCIe device, the kernel module creates the IOMMU page tables for the allocated memory. For CXL devices, the kernel module adds the SPID to the SAT by means of the GFD Component Management Command Set. When a release or share is made, the associated entries are also updated.

\section{EVALUATION}



In this position paper, we present our preliminary simulation results to demonstrate the effectiveness of \textit{LMB} for extending SSD DRAM for L2P indexing.
We compare \textit{LMB} with \textit{Ideal} (all mapping table in onboard DRAM) and \textit{DFTL} \cite{gupta2009dftl} schemes. The \textit{LMB} scheme can be further classified as LMB-CXL and LMB-PCIe to evaluate scenarios that different devices access external memory via \textit{LMB}. We also use PCIe Gen4 and Gen5 SSDs (shown in Table \ref{tab:ssd_specs}) to evaluate the impact of different PCIe standards. We evaluate these schemes with FIO \cite{fio} under workloads of rand/seq writes and rand/seq reads. Additionally, we use the libaio IO engine with a queue depth of 64 and an IO size of 4KB.

\textbf{Prototype implementation.} \textit{LMB} and \textit{DFTL} schemes store mapping tables either on CXL memory or on flash. Due to the scarcity of real CXL development boards, we simulate LMB-CXL and LMB-PCIe on a PCIe Gen4 SSD and a PCIe Gen5 SSD with modified firmware, particularly the L2P indexing module. We use two devices mainly to observe the potential impact of CXL extra latency on devices with different baseline performance. Baseline performance variations between the two SSDs result in different simulation outputs under a same condition.

We add latency in the \textit{Ideal} scheme's L2P indexing for \textit{LMB} and \textit{DFTL} simulation. An additional 25$\mu$s (a flash read operation) is added in \textit{Ideal} to simulate \textit{DFTL} cache miss, while 880ns and 1190ns is added to simulate LMB-PCIe on PCIe Gen4 and Gen5 SSDs, respectively. A 190ns latency is added to simulate LMB-CXL.


\begin{table}[t]
\footnotesize
\centering
\caption{SSD Specifications}
\vspace{-4mm}
\label{tab:ssd_specs}
\begin{tabular}{p{3cm}|p{1.9cm}|p{1.9cm}}
\hline
\textbf{Parameter} & \textbf{PCIe Gen4 x4} & \textbf{PCIe Gen5 x4} \\
\hline
\hline

Capacity (TB) & 7.68 & 7.68 \\
\hline
4KB Rand R/W IOPS & 1750/340 & 2800/700 \\
\hline
128KB Seq R/W BW (GB/s) & 7.2/6.8 & 14/10 \\
\hline
4KB Rand R/W Lat.($\mu$s) & 67/9 & 56/8 \\
\hline
\end{tabular}
\vspace{-0mm}
\end{table}

\begin{figure}[t]
    \centering
    \begin{subfigure}[h]{0.4\textwidth}
        {\includegraphics[width=\linewidth]{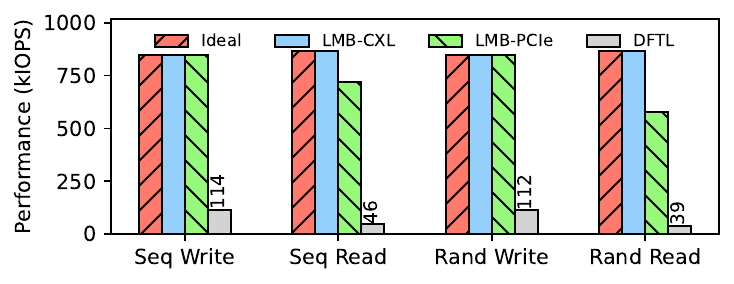}}
        \caption{PCIe Gen4 SSD.}
        \label{fig:gen4}
    \end{subfigure}
    \begin{subfigure}[h]{0.4\textwidth}
        {\includegraphics[width=\linewidth]{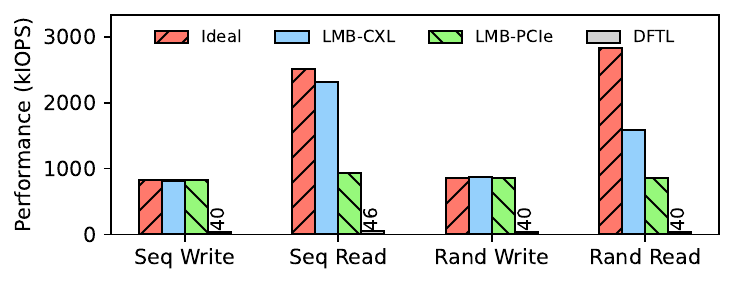}}
        \caption{PCIe Gen5 SSD.}
        \label{fig:gen5}
    \end{subfigure}
    \vspace{-4mm}
    \caption{Comparison of LMB scheme with Ideal (DRAM) and DFTL in the context of PCIe SSD L2P index scenarios.}
    \vspace{-4mm}
    \label{fig:l2p}
\end{figure}

\subsection{Evaluation on Commercial SSDs}

\subsubsection{PCIe Gen4 SSD Evaluation}
We first evaluate the performance of the LMB on a PCIe Gen4 SSD, which is the mainstream product in current SSD market. 

Figure \ref{fig:l2p}(a) illustrates the performance of different schemes. 
\textbf{For write workloads}, LMB-CXL and LMB-PCIe match the Ideal scheme's throughput, outperforming DFTL by $7 \times$. \textbf{For read workloads}, LMB-CXL maintains similar Ideal scheme performance, yet LMB-PCIe experiences a 16.6\% sequential and 13.3\% random read performance drop. Still, it surpasses DFTL by $14 \times$. Given the superior performance of the TLC SSD over the QLC SSD, we posit that the LMB mechanism will effectively handle L2P indexing on high-capacity QLC SSDs to resolve their onboard memory shortage issue.



\subsubsection{PCIe Gen5 SSD Evaluation}
Figure \ref{fig:l2p}(b) compares the performance of different schemes on a PCIe Gen5 SSD. \textbf{For write workloads}, both LMB-CXL and LMB-PCIe deliver similar performance to the \textit{Ideal} scheme, which can achieve $20 \times$ higher throughput than the \textit{DFTL} scheme. \textbf{For read workloads}, both LMB-CXL and LMB-PCIe show performance degradation. More specifically, LMB-CXL achieves 8\% and 56\% lower throughput than the \textit{Ideal} scheme for sequential and random reads, respectively. Meanwhile, LMB-PCIe shows a more severe performance degradation. It shows 62\% and 70\% lower throughput than the \textit{Ideal} scheme for sequential and random reads. Despite that, LMB-PCIe can outperform the \textit{DFTL} scheme by $20 \times$.

Test results indicate that introducing hundreds of nanoseconds of additional CXL latency in the indexing significantly impacts high-performance SSD performance. However, these results assume that all indexing are supported by CXL extended memory. By exploiting the locality of actual workloads where most indices hit on-board memory, the impact on device performance by the CXL secondary index will be considerably dismissed. 





\section{CONCLUSION}
We present a unified LMB framework designed to address the issue of onboard DRAM shortages for PCIe devices. This framework exposes the DRAM resource of the CXL memory expander, providing a consolidated memory management mechanism for both CXL and PCIe devices. Upon investigating this issue in real SSDs, it is evident that our LMB solution effectively supplements onboard DRAM and ensures high performance. The rigorous design and implementation of the LMB framework will tackle various challenges, which we will pursue in future work.

\bibliographystyle{plain}
\bibliography{ref}
\end{document}